\documentclass{caosp302}
\usepackage{float}
\usepackage{graphicx}

\articleNo{123}
\pubyear{2007}
\volume{35}
\volnumber{3}
\firstpage{1}
\received{October, 2007}
\accepted{August 28, 2005}

\begin{document}

%
\hauthor{C.\,Cowley, S.\,Hubrig and F.\,Castelli}

\title{Isotopic Anomalies in CP Stars: Helium, Mercury,
Platinum, and Calcium}


%
\author{
        C. R.\,Cowley\inst{1}
      \and 
        S.\,Hubrig \inst{2}
      \and 
        F.\,Castelli\inst{3}
       }

%
\institute{
           Department of Astronomy, University of Michigan,   \\
         Ann Arbor, MI 48109-1042, USA, \email{cowley@umich.edu}
         \and 
           ESO, Casilla 19001, Santiago 19, Chile \email{shubrig@eso.org}
         \and 
           INAF-Osservatorio Astronomico di Trieste, Via G. B. \\
          Tiepolo 11, 34131, Trieste, Italy \email{castelli@oats.inaf.it}
          }

\date{October 8, 2007}

\maketitle

\begin{abstract}
We review the classical observational results for isotopic
abundance variations for several elements in CP stars.  We
concentrate on the ``newest'' anomaly, in calcium.  The
cosmically very rare isotope, $^{48}$Ca can rival and even
dominate the more common, alpha nuclide, $^{40}$Ca.  Relevant
examples are found in the hot, non-magnetic HgMn stars,
and the field horizontal-branch star, Feige 86.   The
calcium anomaly is also present in cool, magnetic stars,
including the notorious HD 101065, Przybylski's star.
\end{abstract}

%
\section{Introduction}
\label{intr}

The earliest observations of isotopic effects in stellar
spectra were for diatomic molecules.
The Michigan
astronomer W. C. Rufus (1916) noted systematic absorptions
in the spectra of R stars that we now know were due to
differing isotopic species in the Swan bands of C$_2$.
At the time,
the Swan bands were known to be due
to carbon, but the identification with the C$_2$ molecule
was made some ten years later (see Sanford 1932).
By the late 1940's, Andrew McKellar (1948)
was able to make a systematic survey of the $^{12}$C/$^{13}$C
ratio in carbon stars by a study of the molecular bands.

The deuterium analog of the famous 21 cm radio line of
neutral hydrogen was observed in the 1960's (Weinreb 1962).
The most extensive observations of isotopic spectra
are for interstellar molecules.  The field was already mature
several decades ago (cf. Wanneir 1980; Wilson and Rood 1994).

\section{$^3$He in stellar spectra}

Investigations of atomic lines for possible isotopic variations
also date from in the middle of the 20th Century.
Greenstein (1951) sought evidence for $^3$He in the solar
spectrum, using laboratory wavelengths of
Fred, et al. (1951), while the Burbidges (1956)
sought
the same isotope in the magnetic CP star, 21 Aql.

The first secure identification of $^3$He in a stellar
spectrum was made by Sargent
and Jugaku (1961).  These authors
found $^3$He in the spectrum of 3 Centauri A.  Hartoog
and Cowley (1979) studied a number of stars for which
$^3$He had been reported.  They listed 8 stars where
the presence of
$^3$He was definite, and another three probable cases.
Among the field stars, $^3$He is rare, and cases with
virtually pure $^3$He are rarer still.

\begin{figure}
\centerline{\includegraphics[width=5.5cm,angle=-90,clip=]{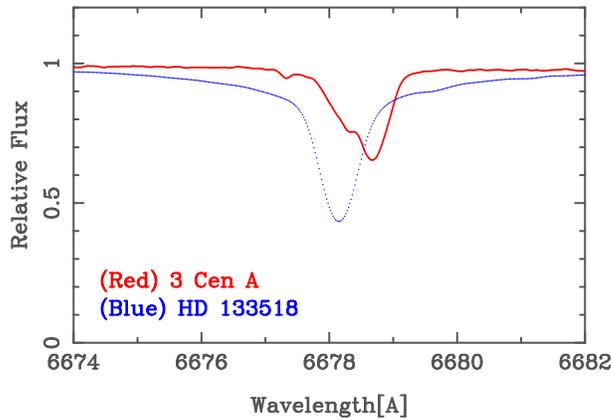}}
\caption{UVES spectra of the $^3$He-star 3 Cen A (solid)
         and HD 133518 (dotted).  Colors appear in online
         version.
         The spectra are from the UVESPOP archive
         (Bagnulo, et al. 2003)}
\label{fig:he3}
\end{figure}

\begin{figure}
\centerline{\includegraphics[width=5.5cm,angle=-90.,clip=]{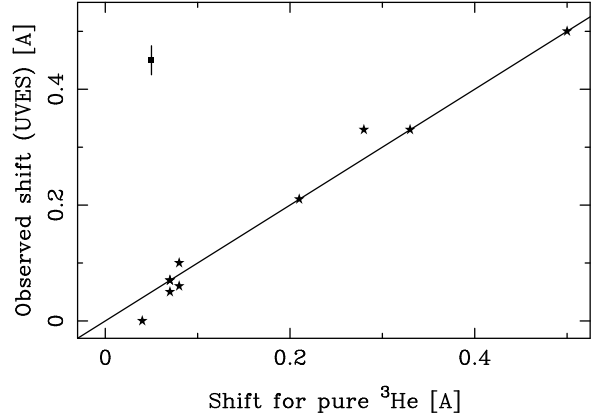}}
\caption{Wavelengths shifts for stellar He I in 3 Cen A
relative to laboratory wavelengths vs. similar shifts for
laboratory $^3$He positions.  The straight line would be
on a 45$^o$ slope, if the
scales of ordinate and abscissa were equal.
The error estimates
$\pm 0.025$ are indicated in the upper left portion of the
plot.  The largest deviations here are probably not due to
measurement errors alone but also to blends or close components.
}
\label{fig:he3sh}
\end{figure}

Fig.~\ref{fig:he3} shows a modern plot of the $\lambda$6876 line
of He I in 3 Cen A and the sharp-lined early B-star HD 133518.
The half-angstrom isotope shift is easily seen in 3 Cen A.
Most of the $^3$He stars show a mixture with $^4$He.  One
cannot, of course, make a credible case for the presence of
an exotic species on the basis of a single wavelength
measurement.  Sargent and Jugaku
compared the wavelengths of 10 He I lines in 3 Cen A and
the B2 IV, MK standard, $\gamma$ Peg.  Fig.~\ref{fig:he3sh}
is a similar plot based on the UVESPOP spectrum, using the same
He I lines as Sargent and Jugaku; we took laboratory
wavelengths rather than stellar as standards.

\begin{figure}[H]
\centerline{\includegraphics[width=5.5cm,angle=90,clip=]{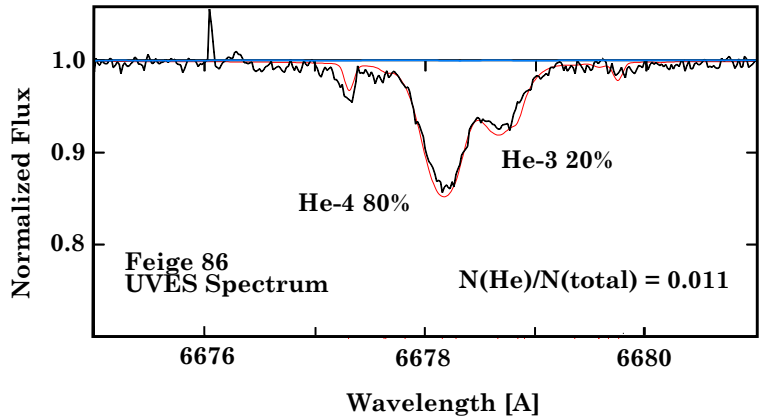}}
\caption{UVES spectrum showing the $\lambda$6678 region in
         the field horizontal branch star Feige 86.  The light,
solid line is the calculated (by FC) spectrum.}
\label{fig:feige2}
\end{figure}

Hartoog (1979) noted
that many of the abundance anomalies 3 Cen A were shared by
the field horizontal branch star Feige 86.  He obtained spectra
of the latter star, and found that it did indeed have an
excess of $^3$He.  A modern spectrum is shown in Fig.~\ref{fig:feige2}.

Heber (1991) discusses the behavior of $^3$He in a number of
horizontal branch stars.  Some
have nearly pure $^3$He.

\section{Isotopic anomalies in mercury and platinum}

W. P. Bidelman (2000) gives a historical account of his identification
of a strong line at $\lambda$3984 as
Hg II in the stars now known as HgMn, or mercury-manganese stars.
Interestingly, it was the spectrum of a related star, 53 Tau,
that greatly strengthened the identification.  In fact, the
laboratory Mn II spectrum
was poorly known, prior to the work of Iglesias and Velasco (1964).
Thus, in 1962, the year of Bidelman's discovery, it seemed possible
the $\lambda$3984 line was due to an unclassified line of
Mn II.  But 53 Tau
showed this could not be, since the 53 Tau spectrum was
replete with strong Mn II lines, but {\it lacked} $\lambda$3984.
There was no plausible candidate for the identification in the
astronomical spectroscopist's Bible, the Multiplet Tables
(Moore 1945).
Bidelman (1962) made the identification with the help of the
MIT Wavelength Tables (Harrison 1939).  But it turns out there
were a number of Hg lines (not all designated as Hg II) near
3984~\AA, and Bidelman learned, in a conversation at a meeting
with the physicist Richard W. Shorthill, that these lines
were due to isotopes of mercury.  This exciting new piece of
information was first published in the popular magazine,
{\it Sky and Telescope} (Federer 1962).

\begin{figure}
\centerline{\includegraphics[width=5.5cm,angle=-90,clip=]{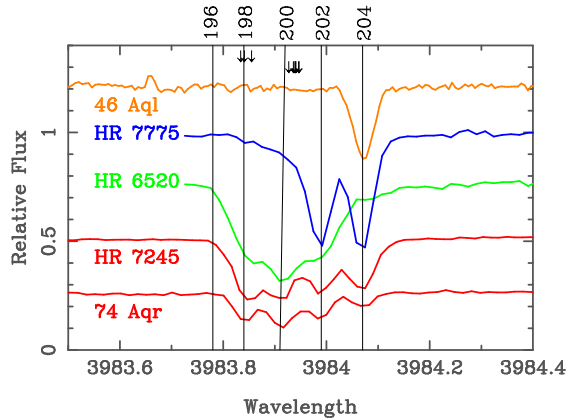}}
\caption{UVES spectra of the $\lambda$3984 region in
5 HgMn stars.  Vertical lines indicate the wavelengths of
the stable, even-A isotopes, indicated above the plot.  Short
arrows indicate the hyperfine components of $^{199}$Hg
and $^{201}$Hg, the stable, odd-A isotopes.}
\label{fig:3984}
\end{figure}

The large shifts of Hg II $\lambda$3984 are caused by nuclear
volume effects.  There are small changes in the volume of nuclei
with differing neutron numbers.  These are sufficient to modify the
electrostatic potential, causing the energy differences that
give rise to the measurable wavelength shifts.  A few examples
are shown in Fig.~\ref{fig:3984}.  Many additional cases,
including the notorious manganese star without $\lambda$3984,
53 Tau, are illustrated in the extensive study by Woolf and
Lambert (1999).

Bidelman (2000) wrote that his confidence in the $\lambda$3984
identification was bolstered by the presence of some Hg I
lines in cooler manganese stars.  Three permitted lines form
Multiplet 1 in Moore's (1945) famous tables.  The transition
is $\rm 6^3P^o - 7^3S$.  The lines decrease in relative
intensity from the strongest, $\lambda$5461, to
the intermediate $\lambda$4358, to
$\lambda$4046, the weakest of the three.  The situation was
not happy, however, because the stellar features did not follow
the laboratory intensities.  The $\lambda$4046 line was too
strong.
In a review of manganese and
related stars, Aller and Ross (1967) referred to $\lambda$4046
as a line ``attributed to Hg I.''  Various speculations
blamed the anomalous intensity on non-LTE or bad oscillator
strengths.

The puzzle over $\lambda$4046 was resolved by Dworetsky (1969),
who identified Pt II in HR 4072, noting the spectrum was also
present in $\iota$ CrB, and $\xi$ Lup.  The strongest of the
Pt II lines has a laboratory wavelength of 4046.45~\AA, so
that it is often closely blended with the Hg I line at
4046.56~\AA.  Both lines are subject to isotope shifts,
already suggested by Dworetsky.  Soon thereafter, Dworetsky
and Vaughan, Jr. (1973) strengthened the case for isotopic
variations in Pt.  Definitive measurements of isotopic shifts
in the laboratory were made by Engleman (1989).

Hubrig, Castelli, and Mathys (1999) were able to determine
specific isotopic abundances of Pt II in three stars.  Their
results are summarized in our Tab.~\ref{tab:pt}, a shortened
version of their Tab.~\ 6.  It shows, as Dworetsky and Vaughan
had pointed out,
the isotopic mixtures were dominated by the two heaviest
stable isotopes--$^{196}$Pt and $^{198}$Pt.

\begin{table}
\small
\begin{center}
\caption{Isotopic abundances of Pt from Hubrig, Castelli, and Mathys
(1999)}
\begin{tabular}{lrrrr}
\hline\hline
Isotope &Terrestrial&HD 35548&HD 141556 &HD 193452 \\
        &abundance &HR1800&$\chi$ Lup& HR 7775 \\
\hline
190  &0.01&0.00 &0.00 &0.00\\
192  &0.79&0.00 &0.00 &0.00\\
194  &32.9&0.00 &0.00 &0.00\\
195  &38.3&0.00 &0.00 &17.5\\
196  &25.2&0.00 &10.00 &55.00\\
198  &7.19&100.0&90.00&27.50 \\
\hline
$\log(N_{\rm Pt}$/$N_{\rm total}$&-10.24$_\odot$&-6.84  &-6.24  &-5.65\\
$\rm [Pt]$                       &              &$+3.40$&$+4.00$&$+4.59$ \\
\hline\hline
\end{tabular}
\end{center}
\label{tab:pt}
\end{table}

\section{Difficult isotopic determinations}

In this section we briefly note two fascinating isotopic
variations.  They share requirement of the best
observational material and the most careful reductions.

\subsection{Thallium in $\chi$ Lup}

Leckrone, et al. (1996) have found evidence for a non-solar
isotope ratio of the two stable thallium isotopes in the
HgMn star $\chi$ Lup.  This was the first star discovered
to have nearly pure $^{204}$Hg, and it was one of several
stars discussed by Dworetsky and Vaughan (1973) as probably
having an excess of the heavy isotopes of platinum.  It is
therefore reasonable that other isotopic anomalies might
manifest themselves in $\chi$ Lup.

Only two lines of Tl II near $\lambda$1909
were identified by Leckrone and
coworkers in the GHRS spectrum of $\chi$ Lup.  Both
stable isotopes, $^{203}$Tl and $^{205}$Tl have nuclear
spin 1/2.  The observed lines
arise in the hyperfine doublet of the intersystem
transition $\rm {6s^2}\, ^1S_0-6s6p\, ^3P_1$.  Because of the
$\rm s^2$ configuration of the lower level, one expects
sensitivity to nuclear volume.  However, the isotopic
shifts were measured in the laboratory, and for both
lines are only about 10 miliangstroms (0.0093 and
0.0105 \AA).  In addition, one of the lines is clearly
blended, so the assertion of an isotopic shift
rests primarily on a single line.

The difficulty of the determination is
further enhanced because
neither isotope dominates, so not even the full
Tl isotopic shifts are measured.
In a tour de force discussion
Leckrone and coworkers nevertheless made a credible case
for a non-standard isotopic ratio.

\subsection{Lithium isotopes in Przybylski's star.}

Most stellar observations of lithium have been made on
the close, D-like, resonance transitions $\rm 2s - 2p$
near 6708~\AA.  There are two stable isotopes, $^6$Li
(spin 1), and $^7$Li (spin 3/2).  However, the hyperfine
splitting is less important than the Zeeman effect,
given a surface field of the order of 3kG.
Approximate wavelengths are shown in Tab.~\ref{tab:li1}
for the 4 D-like transitions: D$_2$ is
$\rm ^2S_{1/2} - ^2P^o_{3/2}$ with (pure LS)
relative intensity
2, while D$_1$ is $\rm ^2S_{1/2} - ^2P^o_{3/2}$
with relative intensity 1.  In the table, the
relative intensities are for
equal isotopic abundances.
Precise wavelengths are given by
Das and Natarajan (2007).

\begin{table}
\small
\begin{center}
\caption{Approximate wavelengths of $\rm 2s - 2p$ transitions
in Li I}
\begin{tabular}{c c c l}
\hline\hline
Isotope &Transition &Int&$\lambda$[\AA] \\ \hline
$^7$Li I &D$_2$ &2& 6707.76 \\
$^7$Li I &D$_1$ &1& 6707.92 \\
$^6$Li I &D$_2$ &2&6707.92 \\
$^6$Li I &D$_1$ &1&6708.07   \\
\hline\hline
\end{tabular}
\end{center}
\label{tab:li1}
\end{table}

When one considers that Zeeman components are split
by 0.1 to 0.2~\AA, {\it and} that the region is also
blended, the difficulty of the determining the relative
contribution of the isotopes is apparent.

Polosukhina and her colleagues (Polosukhina et al. 2004)
described an international project to study lithium in
cool Ap stars.  These workers have published
$^6$Li/$^7$Li
ratios in several stars (cf. Shavrina, et al 2004).

\section{Heavy calcium in HgMn and magnetic Ap stars}

Castelli and Hubrig (2004) announced the identification
of unusual isotopic mixtures of Ca in HgMn stars.  The
determination was possible because of large {\it specific
mass shifts} in the lines of the infrared triplet (IRT).
N\"{o}rtersh\"{a}user, et al. (1998) discussed the
laboratory shifts of all 5 stable isotopes.  The largest
shifts occur for $^{48}$Ca.  To within 5\%, all three
IRT lines of $^{48}$Ca II are shifted by 0.2~\AA\, with
respect to lines from $^{40}$Ca II.  The largest shifts
are easily measurable.

Cowley and Hubrig (2005) found similar shifts of IRT lines
in Ap stars of the magnetic sequence.  Cowley, et al.
(2007, henceforth, CHCGW) discussed Ca isotopic shifts in nearly
70 stars including an Am, and Fm$\delta$Del, an N star
and a weak Barium II star.  Most (non-Ap) stars,
including the sun,
show small shifts with respect to the laboratory positions
for a terrestrial mixture.  The meaning of the small shifts
is unclear, but may be related to the saturation
of stellar lines relative to laboratory measurements.

\begin{figure}[h]
\centerline{\includegraphics[width=5.5cm,angle=-90,clip=]{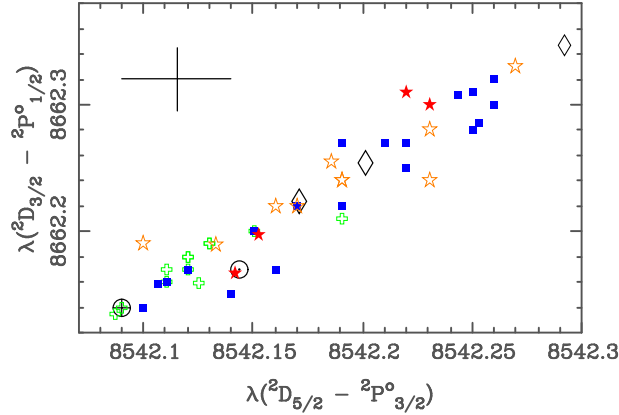}}
\caption{UVES measurements of the strongest lines of the Ca II
IRT.  Terrestrial and solar wavelengths are indicated by
$\oplus$ and $\odot$.  Diamonds indicate positions for pure
isotopes of mass 42, 43, 44, and 48.  Filled squares are HgMn
stars.  Filled and open stars are, respectively, for noAp and
roAp types.  Miscellaneous types are shown with open crosses.
Error estimates are shown to the upper left.}
\label{fig:4262}
\end{figure}

Small displacements of stellar wavelengths are typically
caused by blends.  Blending is highly likely in the complex
spectra of CP stars.  Even with the relatively simple HgMn
spectra, we have noted the unusual coincidence of lines
of Hg I with Pt II.  An indisputable case for isotopic
anomalies requires a demonstration based on a number of
wavelength shifts in agreement with
predictions (cf. Fig.~\ref{fig:he3sh}).
At the time of the initial discovery, many of the ESO
observations missed the strongest line $\lambda$8542
of the IRT because of an order gap in the echelle
spectra.  After reconfiguration of the UVES spectrograph
in November 2004,
it was possible to show that all three lines had
shifts commensurate with heavy calcium isotopes.

Fig.~\ref{fig:4262} shows that the displacements of the
$\lambda$8542 and $\lambda$8662 lines are closely correlated.

In our work, we have assumed it impossible to distinguish
between weighted mixtures of $^{40}$Ca and $^{48}$Ca, and
lines from pure isotopes with intermediate weights.  Ryabchikova
and her colleagues (cf. Ryabchikova 2005,
and also her Review, these proceedings) have made careful
syntheses of the weakest IRT line, $\lambda$8498, and find
evidence not only of vertical stratification, but isotopic
separations.  In those stars with minimal Zeeman broadening,
the core of the $\lambda$8498 line is displaced to the red with
respect to the wings.  This may be explained if the heavy
isotope is pushed to the highest atmospheric levels, while
the deepest layers may have a nearly solar mixture--primarily
$^{40}$Ca.

\begin{figure}
\centerline{\includegraphics[width=5.5cm,angle=-90,clip=]{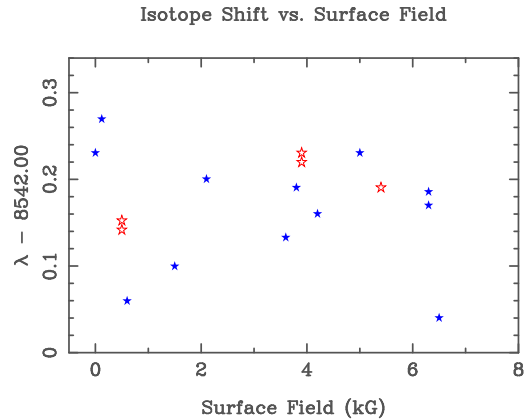}}
\caption{Wavelength shifts of $\lambda$8542 vs. surface field.
Open symbols are for roAp stars; the close pairs are
independent measurements for $\alpha$ Cir and $\gamma$ Equ,
giving some indication of measurement errors.  The lone open
star is for $\beta$ CrB, whose roAp status is relatively new
(Kurtz, Elkin, and Mathys 2007).  A point for HD 154708 would
fall well beyond the upper-right corner.  This plot should be
compared with Fig.~7 of CHCGW.
}
\label{fig:mag42}
\end{figure}

These workers also noted that several stars with surface
fields larger than 5 kG had little or no displacement
of the $\lambda$8498 core.  They suggested this was a general
trend that might be explained if the calcium isotopes were
influenced by light induced drift.
The effect was found independently
by CHCGW who pointed out
two complications with the overall trend.  One star in particular,
HD 154708, has the highest magnetic field of any stars
investigated so far (Hubrig, et al. 2005).  Nevertheless,
its IRT lines show
significant shifts indicating heavy calcium.  Additionally,
a plot of shift of the strongest line, $\lambda$8542 vs.
surface field shows no trend (even if the wildly discordant
point for HD 154708 is omitted).  This is shown in
Fig.~\ref{fig:mag42}

\section{Regularities and relationships of isotopic anomalies}

The intrinsic strength of the IRT makes it possible to observe
isotopic shifts over a wide range of effective temperatures.
Other anomalies are recognized in much narrower domains.
It is unclear to what extent this is due to the physical mechanisms
that produce the anomalies or to selection effects.  Obviously,
for example,
$^3$He-stars could not be found among the CP stars
whose atmospheres are too cool to have
detectable He I lines.

CHCGW looked for correlations of the calcium and mercury
anomalies.  No trend is seen for the HgMn stars
(CHCGW Fig.~4). The Hg II $\lambda$3984 line is surely
present in some magnetic CP stars, but the line spectrum
is too badly blended for credible inferences about
the isotopic composition of mercury.

Preston and colleagues (cf. Preston, et al.
1971) noted a general tendency for the heavier Hg isotopes
to be found in the coolest HgMn stars.  CHCGW found that
when a wider class of objects was examined, the overall
correlation degenerated.  For example, the hottest star
for which they measured $\lambda$3984, Feige 86, showed
a very large wavelength shift, indicating $^{204}$Hg
(cf. their Fig.~5).

CHCGW measured IRT shifts in stars with temperatures ranging
from 6600 to 13600K.  There is no overall systematic trend
of Ca isotopic shifts with temperature.  Weak correlations
may exist separately, and in the opposite sense for magnetic
and non-magnetic CP stars (cf. their Fig.~6).

{}
\end{document}